# Securing Accelerators with Dynamic Information Flow Tracking

**Luca Piccolboni**
Columbia University, USA
piccolboni@cs.columbia.edu

**Giuseppe Di Guglielmo**
Columbia University, USA
giuseppe@cs.columbia.edu

**Luca P. Carloni**
Columbia University, USA
luca@cs.columbia.edu

## 1 HARDWARE DEMO OBJECTIVES

Systems-on-chip (SoCs) are becoming heterogeneous: they combine general-purpose processor cores with application-specific hardware components, also known as accelerators, to improve performance and energy efficiency. The advantages of heterogeneity, however, come at a price of threatening security. The architectural dissimilarities of processors and accelerators require revisiting the current security techniques. With this hardware demo, we show how accelerators can break dynamic information flow tracking (DIFT) [1], a well-known security technique that protects systems against software-based attacks. We also describe how the security guarantees of DIFT can be re-established with a hardware solution that has low performance and area penalties [2].

## 2 INTRODUCTION

**Dynamic Information Flow Tracking**

DIFT [1] is a security technique that prevents a variety of software-based attacks. The key idea is to use *tags* to mark data. For example, in the context of privacy protection the tags are used to mark as sensitive the data that must not be leaked out from an application. DIFT decouples the concepts of policy (what to do) and mechanism (how to do it). The policy defines which data are sensitive and the restrictions to apply on their use. The mechanism ensures that the tags are propagated in the system. We can adopt two schemes to manage the tags [2]. With the coupled scheme, each tag is stored physically with its associated data (same address), i.e., the memory word is extended to include the tag. With the decoupled scheme, the tags are stored separately from the data (different addresses), often in a protected region in memory.

**Target Application and its Vulnerabilities**

We consider an application for an assisted living system as a case study. The application monitors people in a hospital to detect when a patient needs help, e.g., when a patient falls. For privacy protection, the faces of the people, for example those who visit the patients, need to be obfuscated. In this demo, we focus on the part of the application that performs face obfuscation. The application applies a blur filter to a rectangular portion of the input image, called *patch*, defined by four parameters: `i_row_blur`, `e_row_blur`, `i_col_blur` and `e_col_blur`. It produces as output the same image where the pixels inside the patch (sensitive information) are blurred. First, we implemented this function in software and then we designed an accelerator to perform the same computation

This work was supported in part by DARPA SSITH (HR0011-18-C-0017).

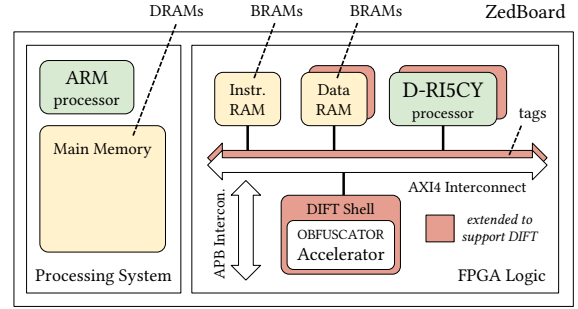

**Fig. 1: Tethered SoC we developed for the demo.**

in hardware to improve performance and energy efficiency (and show why accelerators must be secured with DIFT).

An attacker exploits a *vulnerability* by modifying the configuration parameters of the application in order to reduce the size of the obfuscated area or to move it. This attack may reveal the faces of some people and cause sensitive information to leak out. To protect against this attack, we mark the pixels of the patch as sensitive. All the other pixels are non-sensitive and thus not marked. We enforce a policy that requires all data produced as output by either the software or the hardware implementation must be non-sensitive. The blur filter makes the information non-sensitive and must be applied to all the pixels inside the patch (sensitive area). If an attacker tampers with the sensitive area, we raise an exception before a sensitive pixel could be produced as output.

**Target System-on-Chip Architecture**

The application runs on the tethered SoC shown in Fig. 1. We extended an open-source embedded SoC platform called PULPino [3] to include support for DIFT. We implemented a coupled tagging scheme. We extended the RI5CY processor in PULPino (RV32I ISA) to support tagging and we called it D-RI5CY [4]. We augmented memories and communication channels to accommodate the tags (Fig. 1). In addition, we integrated the accelerator to perform the obfuscation in hardware. The accelerator can be protected with the DIFT shell [2]. In the demo, we show that without the DIFT shell the accelerator can be used as attack vector to compromise the target application, even if the rest of the SoC supports DIFT. We synthesized the SoC platform targeting the reconfigurable logic available on a ZedBoard (Xilinx XC7Z020). After the ARM hard-core processor loads the data into the data memory and the instructions into the instruction memory, the application executes in bare metal on the soft-core D-RI5CY, which optionally invokes the accelerator to blur the image.

**DIFT-Enhanced Hardware Accelerator**

**Accelerator.** We designed the accelerator in SystemC and synthesized it with high-level synthesis [5, 6]. The software first configures the accelerators through a set of registers. The registers are memory mapped and they specify where the input and output images of the accelerator are in memory as well as the values of the four parameters that define the patch boundary. Then, the accelerator performs the obfuscation.

**DIFT Shell.** To support DIFT on the accelerator, we adopted the PAGURUS methodology [2] to design a shell circuit that encloses the accelerator and manages the tags required by DIFT. The accelerator remains unaware of DIFT and does not need to be modified. The shell intercepts the read and write requests of the accelerator to the data memory and it modifies them to include the tags (for tag propagation). The shell also checks that the tags are consistent with the security policy (tag checking), i.e., only the sensitive pixels are obfuscated.

**Demo Objectives and Contributions**

This demo shows the importance of protecting accelerators with DIFT. We show that, even if the entire SoC architecture implements DIFT, a single accelerator that does not support DIFT is sufficient to compromise an application (in this case by leaking sensitive information). We also show the effectiveness of the DIFT shell in protecting the accelerator. The shell has negligible impact on the performance and cost of the accelerator, while it offers strong security guarantees. With respect to other DIFT approaches in the literature, we target heterogeneous SoCs (see the related work reported in [2]).

## 3 ASSUMPTIONS AND ATTACK MODEL

We assume that the hardware is trusted, i.e., no hardware Trojans. The architecture may include some common hardware defenses, e.g., non-executable memory. In this demo we address software-based attacks, e.g., buffer-overflow attacks, etc. We assume that the application has one or more vulnerabilities that permit to modify its configuration parameters. The attacker tries to exploit these vulnerabilities through common I/O interfaces, with the goal of affecting the confidentiality of the hardware-accelerated software application.

## 4 EXPERIMENTAL RESULTS

**D-RI5CY.** We extended the data memory of PULPino (data RAM) from 32KB to 36KB to accommodate the tags. This introduced a 12.5% overhead. The integration of DIFT required an overall increase in the usage of LUT resources that does not exceed 0.82% with respect to the original PULPino SoC. In addition, there is no impact on the processor performance because the tags are processed in parallel and independently from the instruction execution across all pipeline stages [4].

**Accelerator.** We synthesized the accelerator and the shell with Cadence Stratus HLS 17.20 and Xilinx Vivado 2015.1. The shell requires only ~200 LUTs and ~700 flops/latches. The performance overhead on the accelerator execution is negligible (the tags are transferred in parallel with the data).

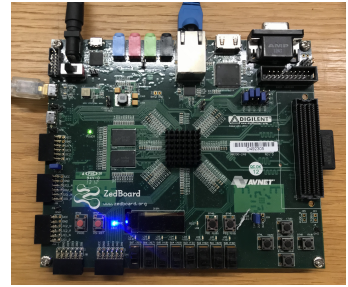

**Fig. 2: The demo runs on a ZedBoard prototyping system.**

## 5 KEY OBSERVATIONS AND OUTCOMES

We prepared a video of the hardware demo that is available at: *http://www.cs.columbia.edu/~piccolboni/demos/host2019.mp4*. The demo runs on the prototyping system shown in Fig. 2. We run the application (Section 2) in four different scenarios:

**Scenario #1:** the software performs the obfuscation (D-RI5CY); the attack and the DIFT support are disabled; this shows the expected execution of the application;

**Scenario #2:** the accelerator performs the obfuscation, the attack and the DIFT support are still disabled; this shows the correctness of the hardware accelerator;

**Scenario #3:** the accelerator performs the obfuscation, the attack is enabled and the accelerator is not protected by the DIFT shell; this scenario shows that the attack succeeds: even if the entire system (except the accelerator) supports DIFT, the application can be compromised by offloading the computation in hardware;

**Scenario #4:** the accelerator performs the obfuscation, the attack is enabled and the accelerator is protected with the shell; this scenario shows that the shell effectively propagates the tags from the processor to the accelerator and vice versa; in this case the attack fails.

Our demo shows why a holistic DIFT approach is needed to prevent software-based attacks. While more convoluted and critical attacks can be implemented, the software-based attack discussed in this demo is representative of the vulnerabilities that can be exploited in heterogeneous SoCs [2].